IAC-10-E3.2.6

# CONTRIBUTIONS OF THE UNITED NATIONS OFFICE FOR OUTER SPACE AFFAIRS TO THE INTERNATIONAL SPACE WEATHER INITIATIVE (ISWI)

H.J. Haubold
United Nations Office for Outer Space Affairs, hans.haubold@unoosa.org

S. Gadimova
United Nations Office for Outer Space Affairs, sharafat.gadimova@unoosa.org

W. Balogh
United Nations Office for Outer Space Affairs, werner.balogh@unoosa.org

ABSTRACT

In 2010, the United Nations Committee on the Peaceful Uses of Outer Space began consideration of a new agenda item under a three-year work plan on the International Space Weather Initiative (ISWI). The main objectives of ISWI are to contribute to the development of the scientific insight necessary to improve understanding and forecasting capabilities of space weather as well as to education and public outreach. The United Nations Programme on Space Applications, implemented by the Office for Outer Space Affairs, is implementing ISWI in the framework of its United Nations Basic Space Science Initiative (UNBSSI), a long-term effort, launched in 1991, for the development of basic space science and for international and regional cooperation in this field on a worldwide basis, particularly in developing countries. UNBSSI encompassed a series of workshops, held from 1991 to 2004, which addressed the status of basic space science in Africa, Asia and the Pacific, Latin America and the Caribbean, and Western Asia. As a result several small astronomical research facilities have been inaugurated and education programmes at the university level were established. Between 2005 and 2009, the UNBSSI activities were dedicated to promoting activities related to the International Heliophysical Year 2007 (IHY), which contributed to the establishment of a series of worldwide ground-based instrument networks, a node of which is also operated by the Office for Outer Space Affairs. Building on these accomplishments, UNBSSI is now focussing on the ISWI.

## I. THE INTERNATIONAL SPACE WEATHER INITIATIVE

Globally there is growing interest in better understanding solar-terrestrial interactions, particularly patterns and trends in space weather. This is not only for scientific reasons, but also because the reliable operation of ground- and space-based assets and infrastructures is increasingly dependent on their robustness against the detrimental effects of space weather. The depictions contained in Figure 1 have been frequently used by the United Nations Office for Outer Space Affairs (UNOOSA) to illustrate the impact of solar activity on planet Earth. Consequently, in February 2009, United Nations Member States represented in the United Nations Committee on the Peaceful Uses of Outer Space (UNCOPUOS) proposed the International Space Weather Initiative (ISWI) as a new agenda item to be taken up in the UNCOPUOS Scientific and Technical Subcommittee under a three-year workplan from 2010 to 2012 [1]:

- 2010 Consider reports on regional and international plans. Encourage continued operation of existing instrument arrays and encourage new instrument deployments;

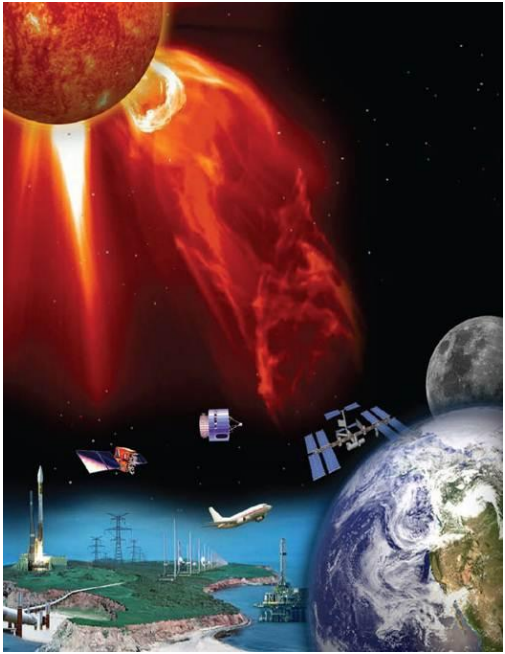

**Fig. 1 The impact of solar activity on planet Earth (NASA)**

- 2011 Consider reports on regional and international plans. Identify gaps and synergies in ongoing activities. Encourage continued

- operation of existing instrument arrays and encourage new instrument deployments;

- 2012 Finalize a report on regional and international plans. Encourage continued operation of existing instrument arrays and encourage new instrument deployments.

The agenda item was subsequently endorsed by the Committee in June 2009 and by the General Assembly in December 2009 [2], [3].

ISWI contributes to the observation of space weather phenomena through the deployment of ground-based instrument arrays (magnetometers, solar telescopes, VLF monitors, GPS receivers, particle detectors) and the sharing of recorded data among researchers around the world. It is implemented by UNOOSA in the framework of its United Nations Basic Space Science Initiative (UNBSSI). Member States are also called upon to raise awareness for international cooperation to address ISWI.

UNBSSI is a long-term effort by UNOOSA for the development of space science and regional and international cooperation in this field on a worldwide basis, particularly in the developing countries. Between 1991 and 2004 a series of basic space science workshops were held to address the status of space science in Asia and the Pacific, Latin America and the Caribbean, Africa, Western Asia and Europe. Based on the recommendation of these workshops, UNBSSI resulted in the inauguration of planetariums and small astronomical telescope facilities in developing countries as well as the development of educational materials for teaching and observing programmes and contributed to improving access to astrophysical data systems and the promotion of the virtual observatory concept for the development of astronomy on a worldwide basis [4], [5].

Following the proclamation of 2007 as the International Heliophysical Year (IHY), proposed by UNCOPUOS and supported by the General Assembly, UNBSSI activities focused on the promotion of activities related to the IHY [6], [7]. Between 2005 and 2009 a series of workshops on basic space science and the International Heliophysical Year 2007 were held [8]. IHY involved thousands of scientists from more than 70 United Nations Member States and included the deployment of a series of instrumentation arrays, including in the developing countries. This result was the outcome of the early recognition in the planning of the IHY that the understanding of solar-terrestrial interaction was limited due to the lack of observations in key geographical areas. The issue was addressed in the workshops and facilitated the collaboration between research scientists in scientifically interesting geographical locations and researchers in countries with the expertise to build the necessary observation instruments. Science teams emerged, grouped around a lead scientist who provided the instruments or fabrication plans for instruments in an instrument array. The nation hosting instruments of an array provided support for the local scientists and the required research infrastructure. All data and data analysis activities are shared within each science team. The 14 instrument array deployed during the IHY included magnetometers to measure the Earth's magnetic field, radio antennas to observer solar coronal mass ejections, GPS receivers, very low frequency radio receivers, all-sky cameras to observer the ionosphere and muon detectors to observer energetic particles [9]. Currently, more than 1,000 instruments are operational in these instrument arrays, which include the Scintillation Network Decision Aid (SCINDA), the Atmospheric Weather Electromagnetic System of Observation, Modeling and Education (AWESOME), the Sudden Ionospheric Disturbances (SID) monitor, the Remote Equatorial Nighttime Observatory for Ionospheric Regions (RENOIR), the Compound Astronomical Low-cost Low-frequency Instrument for Spectroscopy and Transportable Observatory (CALLISTO), the Magnetic Data Acquisition System (MAGDAS) and the African dual frequency GPS network (GPS-Africa).

ISWI is building on the accomplishments made as a result of the celebration of the IHY. The ISWI objectives can be summarized as follows: (a) continuing to deploy new instrumentation, (b) developing data analysis processes, (c) developing predictive models using International Space Weather Initiative data from the instrument arrays to improve scientific knowledge and to enable future space weather prediction services and (d) continuing to promote knowledge of heliophysics through education and public outreach [10]. The Initiative is open to scientists from all countries. They can join either as instrument hosts or as instrument providers. Information on the ground-based worldwide instrument arrays was being distributed through a newsletter being published by the Space Environment Research Centre of Kyushu University, Japan (to subscribe send a blank message to ISWInewsletter-on@mail-list.com) and through the website of the International Space Weather Initiative (www.iswi-secretariat.org).

## II. ISWI IN THE UNITED NATIONS PERMANENT SPACE EXHIBIT

For many years UNOOSA has maintained the United Nations Permanent Space Exhibit located at the Vienna International Centre. The exhibit is one of the highlights of the public tours conducted by the United Nations Information Services (UNIS) at the United Nations Office at Vienna. Each year

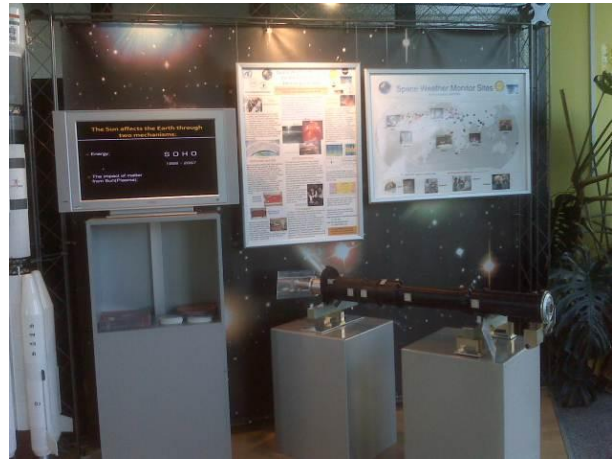

**Fig. 2** ISWI displays in the UNOOSA permanent space exhibit at the United Nations Office at Vienna

several ten-thousands of visitors get to see the exhibit when participating in these tours.

A corner of the exhibit is dedicated to ISWI and displays posters with information on the array of instrument networks deployed during IHY and ISWI. A continuously cycling slide show provides an easy to understand explanation of space weather and its influences on Earth. A model of the COR1 coronograph instrument from the Solar TErrestrial RElations Observatory (STEREO) satellite mission, presented by NASA, is also displayed (see Fig.2). In addition the exhibit also hosts an operational instrument of one of the worldwide ISWI instrument arrays.

## III. INSTRUMENT ARRAY OF SUDDEN IONOSPHERIC DISTURBANCES MONITORS

The Earth's ionosphere is located from approximately 70 km above the Earth's surface and is strongly interacting with the intense X-ray and ultraviolet radiation released by the Sun during solar events. Stanford University has developed the inexpensive Sudden Ionospheric Disturbance monitors (SID) and the more sophisticated Atmospheric Weather Electromagnetic System for Observation Modeling and Education monitors (AWESOME)[*] that can be used to monitor this interaction (see Fig. 3). Students in educational institutions can install and use these instruments at their local high school or university and can thus become part of the worldwide SID or AWSOME instrument array (see Fig. 4).

---

[*] see http://solar-center.stanford.edu/SID

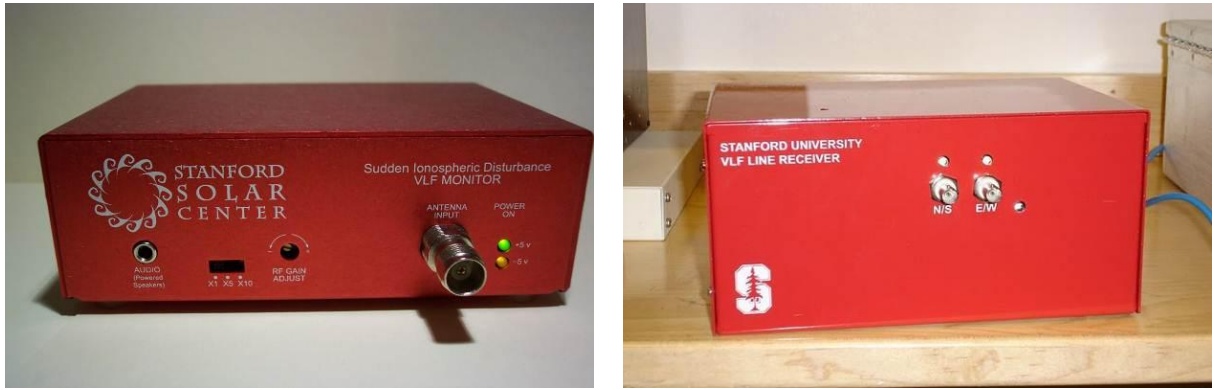

**Fig. 3   SID and AWSOME instruments (Stanford Solar Center)**

With the installation of a SID monitor at the permanent space exhibit at the United Nations Office at Vienna in November 2009, Vienna has become one of the many sites worldwide, reporting occurrences of solar flares as part of the ISWI. The system is simple indeed; the receiver monitors how ionospheric reflections of very low frequency waves in the 15 to 50 kHz range are affected during high energy events such as solar flares. The signals are emitted at transmitter sites designed to communicate with submarines, their measured intensity at the receiver is a direct measure of the polarisation at different levels of the ionosphere. Once the prototype antenna had been placed and connected to the SID receiver provided by Stanford University, UNOOSA started getting results immediately and began automatically transmitting the data from the monitor to Stanford for further analysis on a daily basis. It is indeed a practical and enjoyable contribution to the global cooperation of SID instruments. Since then, a better antenna location has been selected to improve signal quality, and the original SID instrument has been replaced by the next generation receiver, the superSID, enabling UNOOSA to simultaneously monitor stations from all over Europe as well as one station at the east coast of the United States. Multiple sources enable UNOOSA to eliminate noise signals specific to one location, and to verify real ionospheric disturbances through comparison among the data collected from all sites. Experiments with the superSID continue.  Recently, excellent results have been obtained by tuning the antenna to a specific transmitter to eliminate noise at nearby frequencies and by coupling the receiver to the antenna in the most optimal fashion to reduce the load on the antenna circuits.

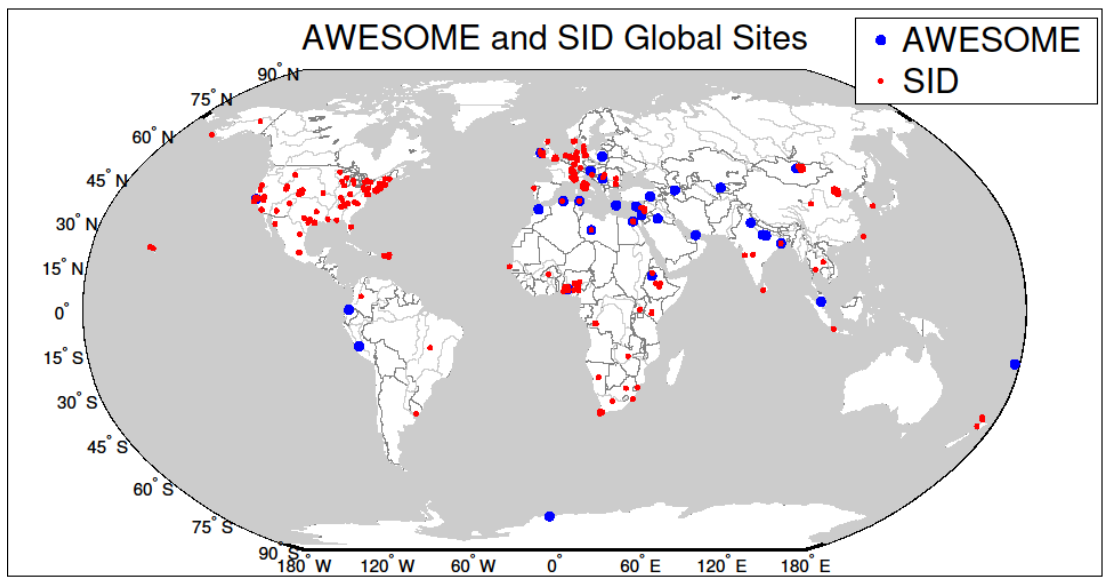

**Fig. 4   World-wide locations of SID and AWSOME instruments**

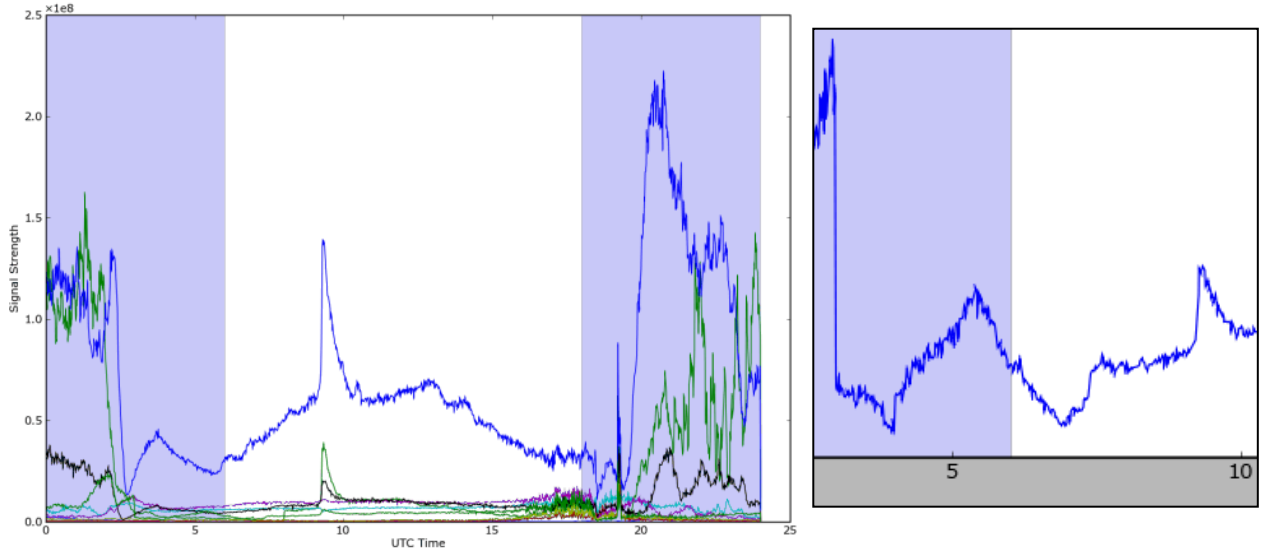

**Fig. 5** Recording of a solar flare at 0900 UCT with magnitude C6 on 12 June 2010 with the SID monitor installed at the United Nations permanent space exhibit at the United Nations Office at Vienna. Graph on the right side depicts detail in the range around 0500 UCT.

The graph shown in Fig. 5 depicts the recording of a flare at 0900 UCT with magnitude C6 on 12 June 2010. Each line on the graph represents the signal amplitude from a specific transmitter. Measured signal strength is displayed over time. During day time the lowest D-layer of the ionosphere, through which the VLF signals have to penetrate to be reflected off the higher located E- and F-layers is weakening the signal. At night time the D-layer disappears and VLF signals are bounced off the E- and F-layers without having to penetrate through the D-layer. During day time in the presence of a Sudden Ionospheric Disturbance (SID) the ionisation of the D-layer is increased such that the VLF signals directly bounce of that layer, which results in the spikes appearing in the measured graph. Solar flares are classified into B-, C-, M-, and X-class flares. While B-class flares are too weak to be detected by SID, and X-class flares saturate the receiver, C- and M-class flares can be readily detected.

The data was confirmed by the Stanford Solar Centre and by another SID site in Germany we correspond with, and is also recorded in the data of the space weather instruments on the United States Geostationary Operational Environmental Satellite (GOES)[†]:

| Event | Beg | Max | End | Obs | Q | Typ | Loc/f | Particulars | Reg |
|---|---|---|---|---|---|---|---|---|---|
| 4370+ | 0902 | 0917 | 0922 | G14 | 5 | XRA | 1-8A | C6.1 3.0E-03 | 1081 |

With a bit of imagination, one could also report a weaker flare noticeable on the NAA transmitter graph as depicted in the detail in Fig. 5.

| Event | Beg | Max | End | Obs | Q | Typ | Loc/f | Particulars | Reg |
|---|---|---|---|---|---|---|---|---|---|
| 4350+ | 0357 | 0406 | 0417 | G14 | 5 | XRA | 1-8A | C1.0 8.3E-04 | 1080 |

The dip at around 0400 UCT corresponds to the GOES data as well, and being of smaller magnitude (C1) is plausible. Note that the changes in amplitude are not always positive; depending on constructive or destructive interference during the reflection in the ionosphere the signal is increased or decreased. While the sensitivity of the SID should be insufficient to pick up B-class flares, the sudden decrease at 0300 UCT could correspond to the following B3 event:

| Event | Beg | Max | End | Obs | Q | Typ | Loc/f | Particulars | Reg |
|---|---|---|---|---|---|---|---|---|---|
| 4340+ | 0253 | 0258 | 0303 | G14 | 5 | XRA | 1-8A | B3.8 1.7E-04 | 1080 |

† see http://www.sec.noaa.gov/ftpmenu/indices/events.html

IV. UNITED NATIONS, ISWI AND ICG

International ISWI workshops have been tentatively scheduled to be hosted by Egypt (2010) for Western Asia, Nigeria (2011) for Africa, and Ecuador (2012) for Latin America and the Caribbean. The 2009 UN/ESA/NASA/JAXA Workshop on Basic Space Science and the International Heliophysical Year 2007, held in Daejeon, in the Republic of Korea in 2009, prepared detailed programmes for ISWI[‡] [8]. The first workshop on ISWI organized jointly by the United Nations, NASA and JAXA will be held in Luxor, Egypt, from 6 to 10 November 2010[§].

ISWI is also supported by the programme on GNSS applications implemented by the Office for Outer Space Affairs in its capacity as the Executive Secretariat of the International Committee on Global Navigation Satellite Systems (ICG)[**] [11]. ICG is contributing to and co-sponsoring several of the ISWI activities [12]. For example, the Office for Outer Space Affairs co-organized a workshop in Morocco to establish scientific and instrumental collaboration for observing the consequences of space weather. The workshop, held in Rabat from 18 to 24 November 2009, was hosted by the Mohammed V University at Souissi on behalf of the Government of Morocco. As a result of the workshop, two magnetometers (Magnetic Data Acquisition System (MAGDAS)), two GPS receivers (GPS-Africa and Scintillation Network Decision Aid (SCINDA)) and one radio spectrometer (Compound Astronomical Low-cost Low-frequency Instrument for Spectroscopy and Transportable Observatory (CALLISTO)) will be transferred to Moroccan observational sites.

Ionospheric modeling using Global Positioning System (GPS) data is the focus of extensive efforts within the GPS community (see Fig. 6). The range error caused by ionospheric delay in GPS signals is currently the largest component that affects the accuracy of positioning and navigation determination using single frequency GPS measurements. Ionospheric modeling is an effective approach for correcting the ionospheric range error and improving the GPS positioning accuracy. The abundance of GPS measurements from worldwide-distributed GPS reference networks, which provide 24-h uninterrupted operational services to record dual-frequency GPS measurements, provides an ideal data source for ionospheric modeling research.

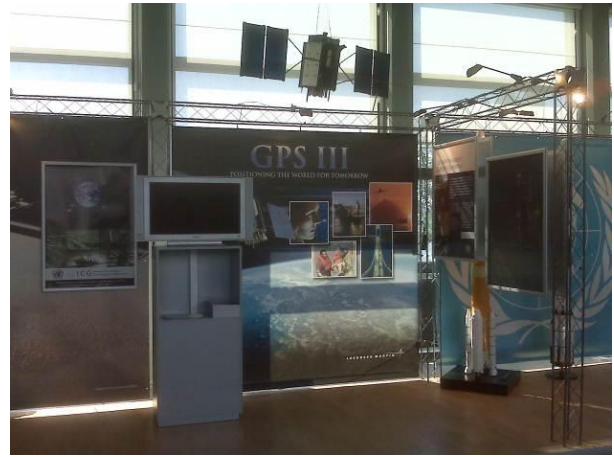

**Fig. 6 A model of a GPS Block III satellite in the United Nations Permanent Space Exhibit**

In the past decade, a large number of GPS reference networks have been deployed worldwide. Therefore, the GPS network facilities provide ionospheric model researchers an excellent data source, allowing the researchers to test, analyze and validate their ionospheric models with extensive GPS data sets. Many efforts of ionospheric model studies have been invested in developing innovative mathematical approaches, to produce better modeling performance and to generate near real-time ionospheric updates.

UNOOSA is planning to operate a GPS receiver for signals from different GNSS to supplement the qualitative results from the SID monitors. The receiver, once operational, would become part of the Scintillation Network Decision Aid (SCINDA)[††] (United States) or GPS-Africa (France).

Furthermore the Basic Space Technology Initiative implemented in the framework of the United Nations Programme on Space Applications will consider the utility of nano- and micro-satellites for the observation of space weather [13].

---

‡ see http://bssihy.kasi.re.kr/unbssw_newsletter.aspx
§ see http://iswi.cu.edu.eg/
** see http://www.icgsecretariat.org

†† see http://www.fas.org/spp/military/program/nssrm/initiatives/scinda.htm

The United Nations/Austria/ESA Symposium on "Small Satellite Programmes for Sustainable Development: Payloads for Small Satellite Programmes", to be held in Graz, Austria, from 21 to 24 September 2010 will include discussions and presentations on this topic[‡‡].

## V. CONCLUSIONS

The United Nations Office for Outer Space Affairs, through the United Nations Programme on Space Applications and in its function as the Executive Secretariat of the International Committee on Global Navigation Satellite Systems will continue to contribute to the implementation of the International Space Weather Initiative.

The Sudden Ionospheric Disturbance (SID) monitor, successfully operating at the United Nations Office at Vienna and will shortly be extended to an Atmospheric Weather Electromagnetic System for Observation Modeling and Education (AWESOME) instrument that provides both solar and night-time research-quality data.

The Office will continue to report its ISWI-related activities to the United Nations Committee on the Peaceful Uses of Outer Space at its forthcoming sessions in 2011.

## REFERENCES

[1] United Nations, "Report of the Scientific and Technical Subcommittee on its forty-sixth session, held in Vienna from 9 to 20 February 2009", A/AC.105/933, 6 March 2009

[2] United Nations, "Report of the Committee on the Peaceful Uses of Outer Space", General Assembly Official Records, Sixty-fourth Session, Supplement No. 20, United Nations, New York, A/64/20, 2009

[3] General Assembly Resolution 64/86, "International cooperation in the peaceful uses of outer space", 10 December 2009

[4] Haubold, H. and Balogh, W., "The United Nations Basic Space Science Initiative (UNBSSI)", Advances in Space Research 43, Elsevier, pp. 1854-1862, 15 June 2009

[5] Haubold, H. and Gadimova, S, "Progress in basic space science education and research: The UNBSSI", Space Policy 26, Elsevier pp. 61-63, 2010

[6] United Nations, "Report of the Committee on the Peaceful Uses of Outer Space", General Assembly Official Records, Sixty-fourth Session, Supplement No. 20, United Nations, New York, A/59/20, para. 135, 2004

[7] General Assembly Resolution 59/116, "International cooperation in the peaceful uses of outer space", 10 December 2004

[8] United Nations Committee on the Peaceful Uses of Outer Space, "Report on the Fifth United Nations/European Space Agency/National Aeronautics and Space Administration/Japan Aerospace Exploration Agency Workshop on Basic Space Science and the International Heliophyiscal Year 2007 (Daejeon, Republic of Korea, 21-25 September 2009)", A/AC.105/964, 19 November 2009

[9] United Nations Office for Outer Space Affairs, "Putting the "I" in the IHY - Comprehensive overview on the worldwide organization of the International Heliophysical Year 2007", http://www.unoosa.org/pdf/publications/ihybookletE.pdf

[10] United Nations Committee on the Peaceful Uses of Outer Space, "Reports on national and regional activities related to the International Space Weather Initiative", A/AC.105/967, 3 December 2009

[11] Gadimova, S. and Haubold, H.J., "The International Committee on Global Navigation Satellite Systems", ICES Yearbook & Directory of Members 2010, pp. 49-56

[12] United Nations Committee on the Peaceful Uses of Outer Space, "Activities carried out in 2009 in the framework of the workplan of the International Committee on Global Navigation Satellite Systems", A/AC.105/950, 11 December 2009

---

[‡‡] see http://www.unoosa.org/oosa/en/SAP/bsti/index.html and http://www.unoosa.org/oosa/en/SAP/act2010/graz/index.html

[13] Balogh, W. and Haubold, H., Proposal for a United Nations Basic Space Technology Initiative, Advances in Space Research 43, Elsevier, pp. 1847-1853, 15 June 2009

*The authors would like to acknowledge the contributions of Mr. Romain Kieffer, Chief of the General Support Section at the United Nations Office at Vienna (UNOV) and his staff for setting-up and maintaining the SID monitor at UNOV.*

*Note: United Nations documents quoted in this paper are available from the website of the Office for Outer Space Affairs at www.unoosa.org and from the Official Document System of the United Nations at documents.un.org.*

*Disclaimer: The views expressed in this paper are purely those of the author and do not necessarily reflect the position of the United Nations and its Office for Outer Space Affairs.*